# Comment on: "Uncertainty in the Multielemental Quantification by Total-Reflection X-ray Fluorescence: Theoretical and Empirical Approximation"

*Leonardo Agüero*[1] *, Leonardo Bennun*[2]

[1] Departamento de Física, Facultad de Ciencias Físicas y Matemáticas, Universidad de Concepción – Chile

[2] Laboratorio de Física Aplicada – Departamento de Física, Facultad de Ciencias Físicas y Matemáticas, Universidad de Concepción – Chile.

Corresponding Author: lbennun@udec.cl

**Abstract**

This work describes an error in the uncertainty assessment of uncertainties of the Total-Reflection X-ray Fluorescence technique Ref [1]. A confusion, between the precision and accuracy of a measurement produced an incomplete evaluation of the uncertainties of the technique.

**1.** Uncertainties and Error Propagation

A mathematical tool used in this article is the propagation of errors formula, given by:

$$u_c[C_y(x_i)] = \sqrt{\sum_i \left(\frac{\partial C_y}{\partial x_i}\right)^2 u^2(x_i)}$$

(1)

with $u_c$ being the combined uncertainty associated with $C_y$, and $u(x_i)$ the standard uncertainty of each one of the independent input variables $x_i$.

## 2. Errors propagation in Total-Reflection X-ray Fluorescence analysis.

The Total Reflection X-ray Fluorescence technique (TXRF) is generally used in conditions in which a linear quantification can be applied, meaning enhancement or self-absorption effects are not significant. Thus, the measured intensity, as a function of the concentration, is expressed as:

$$N_y = C_y S_y$$

(2)

with $N_y$ the intensity of fluorescent radiation of the *y* element, $S_y$ its sensitivity, and $C_y$ the concentration of such an element in the sample. In TXRF analysis, for quantification purposes, a new element is added to the sample. This reference element should be added in an extremely well-known concentration. The intensity of the reference element (*ref*) would be:

$$N_{ref} = C_{ref}\, S_{ref} \qquad (3)$$

The relationship between the intensities indicated by Eqs. (2) and (3) produces a relation between the empirical input parameters. Solving for $C_y$ :

$$C_y \left( C_{ref}, N_y, S_{ref}, N_{ref}, S_y \right) = C_{ref}\, \frac{N_y\, S_{ref}}{N_{ref}\, S_y} \qquad (4)$$

and applying Eq. (1) to Eq. (4), we obtain:

$$u_{c,rel}^2(C_y) = u_{rel}^2(C_{ref}) + u_{rel}^2(N_y) + u_{rel}^2(N_{ref}) + u_{rel}^2(S_{ref}) + u_{rel}^2(S_y) \qquad (5)$$

Eq. (5) evaluates the uncertainty associated with the parameter $C_y$ , depending on every one of the input parameters. This evaluation determines the precision of this technique.

The contribution due to the sensitivities of the uncertainties $u_{rel}^2(S_{ref})$ and $u_{rel}^2(S_y)$ in the TXRF method can be obtained by means of two empirical processes: Model A , based on the direct use of the TXRF functional relations, and Model B, based on linear regression. In the following we will discuss Model A.

2.1 Model based on the use of the TXRF functional relation

Considering eq. (4), we have:

$$S_y = S'_{ref}\, \frac{N'_y\, C'_{ref}}{N'_{ref}\, C'_y} \qquad (6)$$

By applying the same procedure to Eq. (6) as to Eq. (4), the relative uncertainties of the sensitivities for the *y* element are obtained:

$$\left(\frac{u(S_y)}{S_y}\right)^2 = \left(\frac{u(C'_y)}{C'_y}\right)^2 + \left(\frac{u(C'_{ref})}{C'_{ref}}\right)^2 + \left(\frac{u(N'_y)}{N'_y}\right)^2 + \left(\frac{u(N'_{ref})}{N'_{ref}}\right)^2 + \left(\frac{u(S'_{ref})}{S'_{ref}}\right)^2 \qquad (7)$$

The first term of Eq. (7) is the relative uncertainty corresponding to the concentration of the $y$ element in the sample; the second corresponds to the reference element concentrations (*ref*); the third to the intensity of the $y$ element; the fourth to the intensity of the reference element (*ref*); finally the fifth term, that is to say $\left(\frac{u(S'_{ref})}{S'_{ref}}\right)^2$, corresponds to the reference element sensitivity.

If a dissolution is prepared such that the concentration of the specific p element *(pattern)* and the normalization element *ref* are known, for example using certified monoelemental pattern, the sensitivities of the *p* element can be easily obtained, from:

$$\left(\frac{u(S_p)}{S_p}\right)^2 = \left(\frac{u(C'_p)}{C'_p}\right)^2 + \left(\frac{u(C'_{ref})}{C'_{ref}}\right)^2 + \left(\frac{u(N'_p)}{N'_p}\right)^2 + \left(\frac{u(N'_{ref})}{N'_{ref}}\right)^2 + \left(\frac{u(S'_{ref})}{S'_{ref}}\right)^2 \qquad (8)$$

Terms in Eq. (8) correspond to the equivalent descriptions of Eq. (7), but applied to the case of the p element in the sample.

In Ref [1] at Eqs.(7) and (8) the uncertainty values corresponding to the fifth term are not included because $S_{ref}$ has a fixed value (defined to be 1) and, in an implicit way, its uncertainty $u(S'_{ref})$ is considered to be zero. Thus, in Ref [1] it is proposed that:

$$u(S'_{ref}) = 0 \rightarrow \left(\frac{u(S_{ref})}{S_{ref}}\right)^2 = 0 \qquad (9)$$

However, we know that in every experiment the measurement processes are not perfect, so each result has an associated uncertainty. In the case of the sensitivity it can be expressed by:

$$S'_{ref} \pm u(S'_{ref}) \qquad (10)$$

We observe that, when performing this kind of experiment the term $S'_{ref}$ is perfectly known, being is equal to 1. However ,this kownledge does not give us the assurance that the uncertainty of its measurement is zero. Therefore, the necessary error propagation for the TXRF functional relation must be corrected for elements $y$ and $p$, as follows:

$$u_{rel}[S_y] = \frac{u[S_y(x_i)]}{S_y} = \sqrt{\left(\frac{u(C'_y)}{C'_y}\right)^2 + \left(\frac{u(C'_{ref})}{C'_{ref}}\right)^2 + \left(\frac{u(N'_y)}{N'_y}\right)^2 + \left(\frac{u(N'_{ref})}{N'_{ref}}\right)^2 + \left(\frac{u(S'_{ref})}{S'_{ref}}\right)^2}$$

(11)

$$u_{rel}[S_p] = \frac{u[S_p(x_i)]}{S_p} = \sqrt{\left(\frac{u(C'_p)}{C'_p}\right)^2 + \left(\frac{u(C'_{ref})}{C'_{ref}}\right)^2 + \left(\frac{u(N'_p)}{N'_p}\right)^2 + \left(\frac{u(N'_{ref})}{N'_{ref}}\right)^2 + \left(\frac{u(S'_{ref})}{S'_{ref}}\right)^2}$$

(12)

## 3. Conclusions

We have shown that, in an error propagation process, all factors that intervene in a measurement need to be carefully considered, because an error in their interpretation may produce an incorrect result. In the use of the TXRF functional relation, we know that the sensitivity value $S'_{ref}$ is 1, but when a measurement of its value is made, every result is affected by an uncertainty, and a $u(S'_{ref})$ value is found, as is shown by Eq. (10).

The fact that this uncertainty is of scalar nature indicates that its evaluation, by any method which takes into account a complete representation of the system, will produce essentially the same result. That is, if each one of the underlying processes in the studied spectroscopic technique are indeed well described by the chosen models, the results should not differ significantly. In this regard, the Model B discussed in Ref [1] can not be better than the Model A. Once the complete evaluation of uncertainties is made, according to Eqs. (11) and (12), both methods will produce very close results. Due to the original paper being widely cited by the TXRF community, it is very important that a reliable evaluation of the uncertatinties of the technique is developed.